%% file: main_v8_submitted.tex
\documentclass[twocolumn,preprintnumbers,notitlepage,amsmath,amssymb,superscriptaddress,numbers,sort&compress]{revtex4-2}

\usepackage[utf8]{inputenc}
\usepackage[english]{babel}
\usepackage{graphicx,color}
\usepackage{caption}
\usepackage{subcaption}
\usepackage{amsmath,amssymb}
\usepackage{epstopdf}
\usepackage{soul}
\usepackage{tabularx}
\usepackage{tikz}
\usetikzlibrary{backgrounds}
\usetikzlibrary{decorations.pathreplacing}
\usepackage{xcolor}
\usepackage[percent]{overpic}
\usepackage{hyperref}
\usepackage[normalem]{ulem}

\newcommand{\Sim}[1]{\mathrel{\mathop{\kern0pt\sim}\limits_{#1}}}
\newcommand{\To}[1]{\mathrel{\mathop{\kern0pt\to}\limits_{#1}}}
\newcommand{\Propto}[1]{\mathrel{\mathop{\kern0pt\propto}\limits_{#1}}}
\newcommand{\p}[1]{\mathbb{P}\left({#1}\right)}

\newcommand{\dw}{d_\text{w}}

\newcommand{\BESQ}[1]{\text{BESQ}^{#1}}
\newcommand{\BESQabs}[1]{\Tilde{\text{BESQ}}^{#1}}

\definecolor{redblind}{rgb}{0.8627450980392157, 0.0196078431372549, 0.047058823529411764}
\definecolor{blueblind}{rgb}{0.2627450980392157, 0.49019607843137253, 0.7490196078431373}
\definecolor{orangeblind}{rgb}{0.9450980392156862, 0.5764705882352941, 0.17647058823529413}
\definecolor{greenblind}{rgb}{0.3058823529411765, 0.6980392156862745, 0.396078431372549}

\newcommand{\insertpdfpage}[2]{%
  \onecolumngrid           
  \clearpage
  \pagestyle{empty}         
  \thispagestyle{empty}     
  \null
  \vspace*{-1in}            
  \hspace*{-0.90in}         
  \makebox[\paperwidth][l]{%
    \includegraphics[page=#2,width=\paperwidth,height=\paperheight]{#1}%
  }%
  \clearpage
  \twocolumngrid           
  \pagestyle{plain}        
}

\begin{document}

\title{Exact collective first-passage statistics of $N$ trail-interacting walkers
}

\author{P. Pineau}
\affiliation{Laboratoire Jean Perrin, CNRS/Sorbonne Université, 
 4 Place Jussieu, 75005 Paris, France}
\author{J. Br\'emont}
\affiliation{Collège de France, 3 rue d'Ulm, 75005 Paris, France}

\author{O. B\'enichou}
\affiliation{Laboratoire de Physique Th\'eorique de la Mati\`ere Condens\'ee, CNRS/Sorbonne Université, 
 4 Place Jussieu, 75005 Paris, France}
 \author{R. Voituriez}
 \affiliation{Laboratoire Jean Perrin, CNRS/Sorbonne Université, 
 4 Place Jussieu, 75005 Paris, France}
\affiliation{Laboratoire de Physique Th\'eorique de la Mati\`ere Condens\'ee, CNRS/Sorbonne Université, 
 4 Place Jussieu, 75005 Paris, France}

\date{\today}
\begin{abstract}
Memory encoded in the environment mediates interactions between active agents, from trail-following organisms to synthetic active matter depositing persistent tracks. Although such memory is known to strongly affect transport, its consequences for collective first-passage phenomena remain largely unexplored. Here we study $N$ one-dimensional random walkers interacting through a shared trail field. We characterize the  $k^{\rm th}$ (among $N$) arrival time at a fixed target, and the probability that exactly $k$ walkers in $[0,1]$ reach one boundary before the other. For the broad class of self-interacting walkers with a saturating response to the trail, we derive exact expressions for the corresponding persistence exponents and splitting probabilities. Strikingly, despite the strong history-dependent correlations generated by the common environment,  splitting probabilities are exactly identical whether walkers explore simultaneously or one after another. This invariance breaks down for nonsaturating trail interactions. Our results follow from an exact representation of the collective trail field and establish a framework for first-passage phenomena in systems coupled through persistent environmental memory.

\end{abstract}

\maketitle

Many living systems, from bacteria, eukaryotic cells, to ants, modify over long time scales their environment by depositing cues  (eg non diffusive chemicals) or by inducing local  structural perturbations, which will be called trail hereafter. For agents that are responsive to these same cues, the resulting dynamics can yield rich emerging  behaviours, which 
are exploited by many living systems: for instance, cells can modify their environment to either limit or on the contrary enhance space exploration, and ants can  follow and reinforce trails left by others to recruit conspecifics~\cite{Holldobler1990,Dussutour2004,Morgan2009TrailPO, Chalissery2019AntsSA}.

When trail deposition is irreversible, and the trail content  non diffusive,  the global dynamics is genuinely out of equilibrium, and the trail acts as a memory field that integrates the dynamics of agents over time, leading to intrinsically non Markovian and non stationary dynamics. These features, which stand in contrast to the classical models of auto chemotaxis of the Keller Segel family,  have raised an increasing interest in mathematics, physics and biology. However, most of the literature so far focused on single agents, and in particular
a large body of work has been developed on so called  self-interacting random walks~\cite{Amit1983,Toth1995,Toth_Generalized_RK_96,Toth1998,Toth2002,Dumaz2013,Bremont2024,Bremont2025_exponents,Mareche2026,coghiSelfinteractingProcesses},  where memory effects can lead, for instance, to non monotonic density profiles \cite{Dumaz2013,Bremont2024,Romano2026,maggsNonreversibleMonte},  anomalous diffusion~\cite{Romano2026,Dumaz2013}, or self trapping \cite{fosterReinforcedWalksa}.

The case of interacting trail depositing particles has been much less studied, and 
theoretical models   have been introduced only recently and focused on  hydrodynamic limits, either obtained phenomenologically \cite{Adar2024,Bell2025} or derived from minimal microscopic models \cite{pineau2025collectivedynamicstrailinteractingparticles}. Of note, trail mediated interactions are by construction not amenable to classical instantaneous pairwise interactions; their quantitative characterization and their impact on the collective dynamics of $N$ interacting agents is thus non trivial and remains to be elucidated ; this is the main focus of this work.

 Natural observables to quantify  how efficiently a given agent explores space or finds targets are given by first-passage statistics, such as the survival probability and associated persistence exponent \cite{vanKampen, Hughes1996,brayPersistenceFirstPassage}, which characterizes the time needed to reach a target in infinite space ~\cite{Redner2001} and splitting probabilities, which quantify the competition between distinct targets~\cite{Redner1999}. This class of observables has been instrumental to quantify the efficiency of stochastic search processes in a broad sense, which have applications in various fields and across scales \cite{brayPersistenceFirstPassage,Redner2001}. While recent works obtained exact results for these quantities at the single-particle level~\cite{Dumaz2013,Bremont2024,Bremont2025_exponents,Romano2026} or for several interacting Markovian particles \cite{biroli_first-passage_2026,grabsch_semi-infinite_2024}, corresponding results for $N$ random walkers with trail mediated interactions are still lacking, and are not accessible with the recently developed fluctuating hydrodynamics approaches~\cite{pineau2025collectivedynamicstrailinteractingparticles}.

In this Letter, we investigate the first-passage statistics of $N$ self-interacting random walkers (SIRWs). In  dimension $d=1$, the dynamics of mutually interacting SIRWs can be  defined straightforwardly by extending the classical case of a single random walker, and introducing   jump probabilities $P_t(x\pm1|x)$ from site $x$ to $x\pm1$ at time $t$ \cite{Toth2002}:
\begin{equation}
    P_t(x + 1|x) = \frac{w(L_t(x+1))}{w(L_t(x+1)) + w(L_t(x))} = 1-P_t(x-1|x).
    \label{eq:proba_jump}
\end{equation}
Here $L_t(x)$ is the edge local time, i.e., the cumulative number of crossings of the unoriented edge $\{x-1,x\}$ by {\it any} of the SIRWs up to time $t$, and $w$ is a positive weight function \footnote{A site-reinforced variant of SIRWs can also be defined, for which our results still hold.}. Decreasing $w(n)$ corresponds to effective self-repulsion, while increasing $w(n)$ corresponds to self-attraction. Importantly, walkers share the same local-time field $L_t(x)$, and therefore the same spatial memory: they both self- and mutually interact. We focus on two broad classes of SIRWs \cite{Toth_Generalized_RK_96}, characterized by the large-$n$ asymptotics of $w(n)$.

\begin{figure}[t]
    \centering
    \resizebox{\linewidth}{!}{\input{figures/schema.tikz}}
    \caption{Schematic dynamics of three $\text{SATW}_\phi$ walkers launched from $x_0$. Red dots denote walkers, and blue regions denote sites already visited. In the interior of the visited domain, motion is symmetric (left inset), whereas at the boundary the jump probabilities become asymmetric (right inset). Corresponding values of the probabilities are shown above the arrows. The splitting event $(+^k|x_0)$ (in green) corresponds to the event where $k$ (resp. \ $N-k$) walkers are absorbed at wall 1 (resp. \ $N-k$ at wall 0).~\label{fig:schema}}
\end{figure}
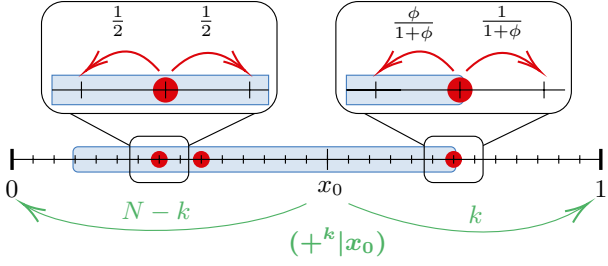

The first class of saturating SIRWs corresponds to a saturating sensitivity to the trail field, $w(n) \xrightarrow[n\to\infty]{} l>0$. Many examples of environment sensing mechanisms at the cell scale are expected to display such saturation, and models of this class have indeed proved to be relevant to describe cell dynamics \cite{BarbierChebbah2022}. Its simplest representative is the once-reinforced walk, denoted SATW$_\phi$, defined by $w(0)=1/\phi$ and $w(n\geq1)=1$ (see Fig.~\ref{fig:schema}); note that $\phi=1$ recovers the case of non-interacting Brownian walkers. More generally, 
all models in this class can be showed to be equivalent asymptotically to  SATW$_\phi$ with $\phi = w(0)^{-1} + \sum_{j=1}^{\infty}\left[w(2j)^{-1}-w(2j-1)^{-1}\right]$, which substantially broadens its scope~\cite{Toth_Generalized_RK_96}. 

The second class of non saturating SIRWs is characterized by $w(n)\to0$, either algebraically (PSRW, $w(n)=n^{-\alpha}$ with $0<\alpha<1$) or exponentially (TSAW, $w(n)=e^{-\beta n}$ with $\beta>0$). These models have been widely studied for a single walker \cite{Dumaz2013,Bremont2025_exponents,Bremont2024,BarbierChebbah2022,coghiSelfinteractingProcesses,kosyginaConvergenceRandom,Romano2026,Toth_Generalized_RK_96} and, more recently, the fluctuating hydrodynamics of interacting walkers of the TSAW class has been derived using stochastic density field theory ~\cite{pineau2025collectivedynamicstrailinteractingparticles}.

In this Letter, we derive analytically exact expressions for  the generalized persistence exponents and splitting probabilities of $N$ SATWs ; our analysis reveals the deep impact of trail mediated interactions on exploration  dynamics, and our conclusions are extended on the basis of numerical simulations to the second class of SIRWs. We first summarize our results below, then provide key steps of their proofs, which are given  in full in Supplementary Material (SM) and last comment on their significance.

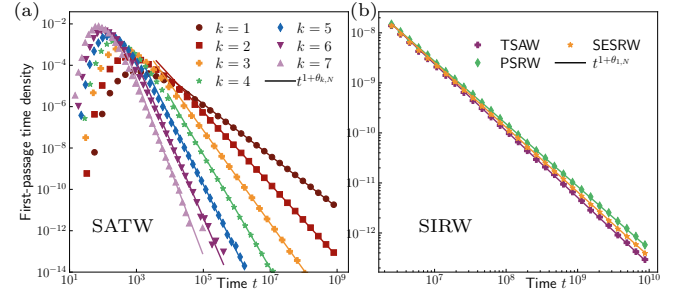
\begin{figure}[ht]
    \centering
    \begin{overpic}[width=\linewidth]{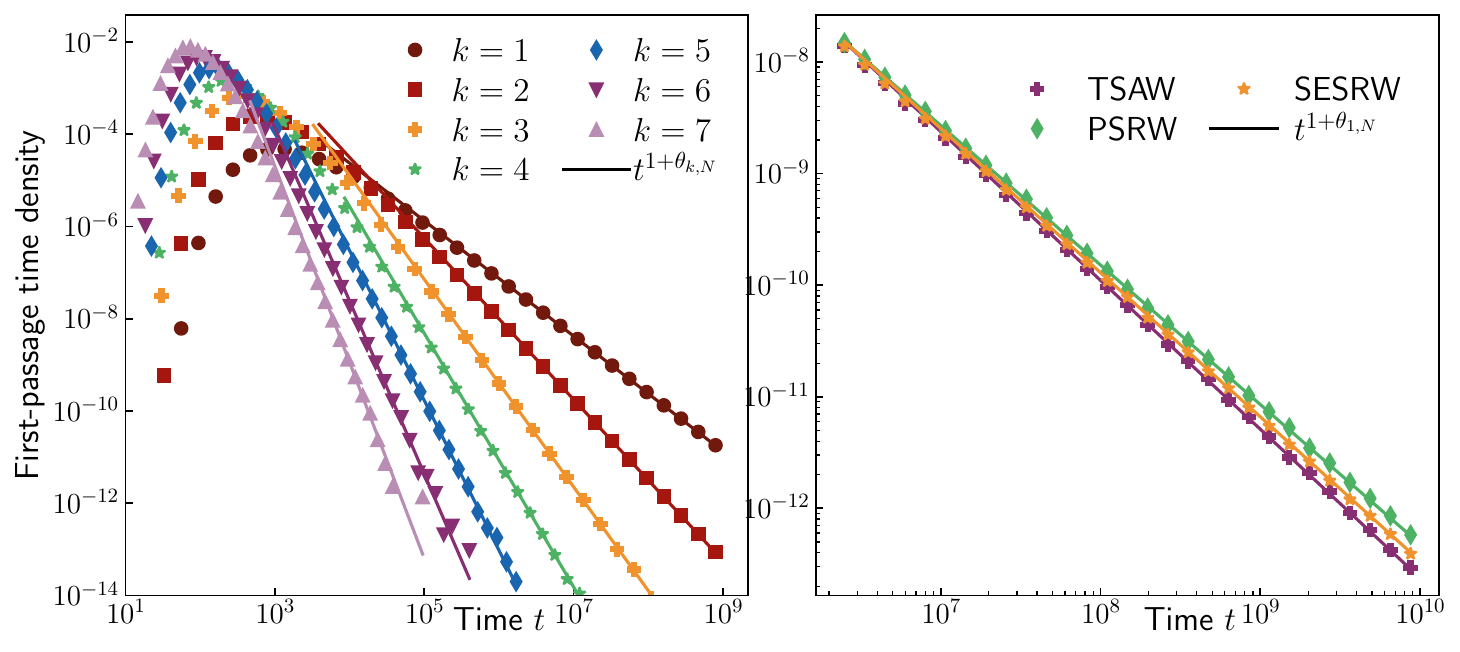}
        \put(0,43){\scriptsize(a)}
        \put(12,10){\scriptsize{SATW}}
        \put(51.5,43){\scriptsize(b)}
        \put(62,10){\scriptsize{SIRW}}
    \end{overpic}
    \caption{Numerical verification multi-particles persistence exponents. First-passage densities of the (a)~$(N-k+1)^{\rm th}$ absorbed SATW$_\phi$ [Eq.~\eqref{eq:theta_kN}], with $N=10$, $\phi=0.5$, and $1\leq k\leq7$, averaged over $10^8$ realizations; (b)~last surviving walker [Eq.~\eqref{eq:theta_1N}] for several SIRW models (TSAW, PSRW, SESRW) with $N=4$, averaged over $10^6$ realizations.  Simulations are performed on a 1D lattice with discrete time and space.\label{fig:panel_fits_FPT}}
\end{figure}

\textit{Results.} The survival probability $S(x_0, t)$ of a $1d$ stochastic process $(X_t)_{t\geq 0}$ is defined in infinite space as the probability that, starting from $x_0 > 0$, the process has not reached $0$ up to time $t$. For symmetric, scale-invariant processes, $S(x_0, t) \propto t^{-\theta}$ at long times, where $\theta$ is the persistence exponent. We generalize this definition to $N$ walkers starting at positions~$\boldsymbol{x_0} = (x_0^1,\dots,x_0^N)$ with $x_0^i>0$ and absorbed at $0$, by introducing $S_{k,N}(\boldsymbol{x_0},t)$, the probability that at least $k$ walkers survive up to time $t$, with associated exponents $\theta_{k,N}$ defined by $S_{k,N}(\boldsymbol{x_0},t)\sim t^{-\theta_{k,N}}$ (the single-particle exponent being $\theta_{1,1}$).  The first main result of this Letter is the exact determination of all such generalized persistent exponents for SATW$_\phi$, and of the last-survival exponent $\theta_{1,N}$ for generic SIRWs:
\begin{align}
        &\theta_{k,N}^{(\text{SATW}_\phi)} = \frac{k+\phi-1}{2}\label{eq:theta_kN}\\ 
        &\theta_{1,N}^{(\text{SIRWs})} = \theta_{1,1}^{(\text{SIRWs})}.\label{eq:theta_1N}
\end{align}  
In particular, these results are consistent with  $\theta_{1,1}^{(\text{SATW}_\phi)} = \phi/2$ obtained in ~\cite{BarbierChebbah2022,Bremont2025_exponents}, and the expected value $\theta_{k,N} = k/2$ obtained straightforwardly for  non-interacting Brownian particles ($\phi=1$). These results are confirmed numerically~\cite{Efron1994,Clauset2009} in Fig.~\ref{fig:panel_fits_FPT} for the SATW$_\phi$, and extended to the full class of saturating SIRWs in~\cite{SM}. These results quantitatively highlight the impact of trail mediated interactions on exploration, where attractive interactions ($\phi>1$) suppress large values for the first-passage times to the target, while repulsive interactions ($\phi<1$) diminish the persistence exponent and favor long, non reactive trajectories. 

\begin{figure}[ht]
    \centering
    \begin{overpic}[width=\linewidth]{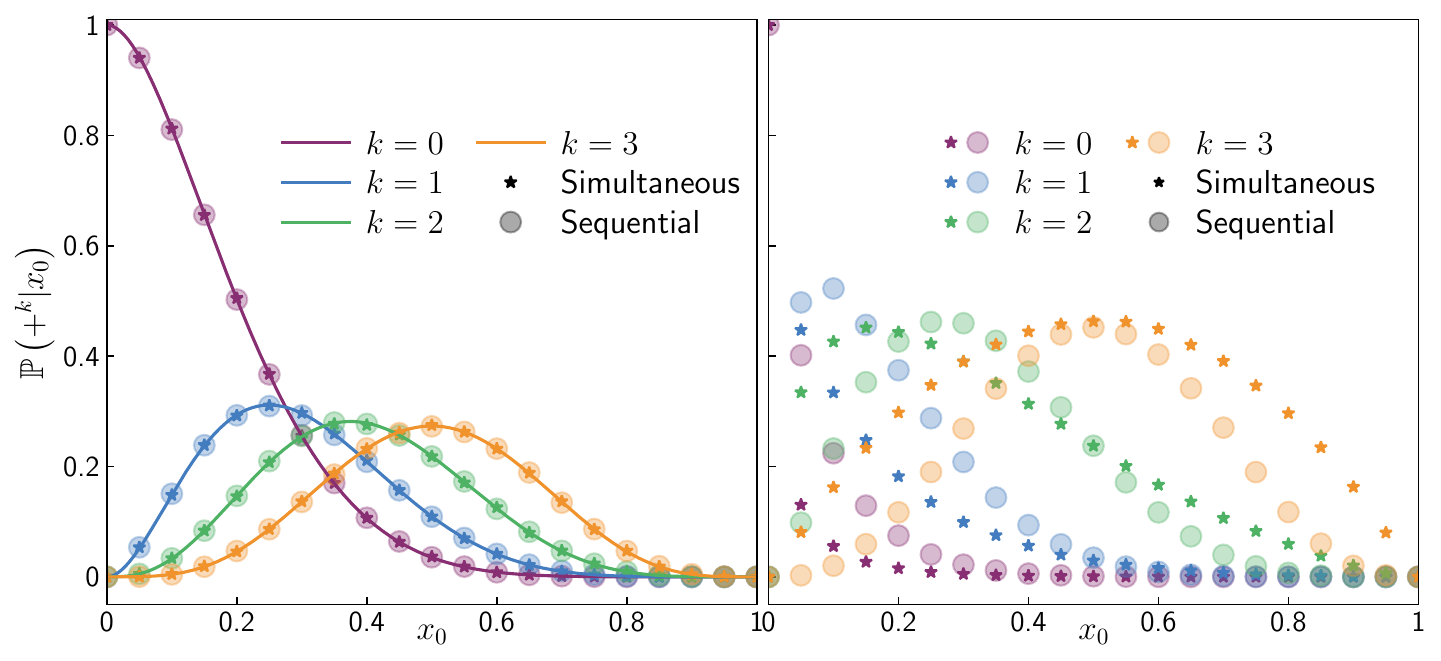}
        \put(13,40){\scriptsize(a): SATW}
        \put(60,40){\scriptsize(b): TSAW}
    \end{overpic}
    \caption{Splitting probabilities for~(a)~$N=6$ SATW$_\phi$ compared with Eqs.~\eqref{eq:modified_binomial} and~\eqref{eq:proba_all_same_wall}, and for~(b)~ $N=6$ TSAW launched from $x_0$, for both simultaneous and sequential protocols. Symbols: simulations averaged over $10^5$ realizations; solid lines: theoretical predictions.\label{fig:panel_splittings}}
\end{figure}

The second quantity of interest is the splitting probability, which, for a single walker, is defined in an  interval bounded by two targets as the probability of absorption by one target rather than the other. In the scaling limit discussed in the following, we choose $[0,1]$ as the interval. The splitting probability is a key observable to quantify competitive processes, and has been showed to be closely related to persistence exponents \cite{Majumdar_hitting_persistence_2010, Levernier2018}. We generalize this definition to $N$ SIRWs, with common origin $0<x_0<1$ and absorbing targets at $0$ and $1$. We denote the event where exactly $k$ walkers are absorbed at target $1$ (with the remaining $N-k$ thus absorbed at wall $0$) as $(+^k|x_0)$.  We consider two different  protocols: sequential start, where walkers explore the domain one after the other, each one starting only after the previous one is absorbed, and simultaneous start, where all walkers start together, which can both correspond to realistic experimental settings (for example \cite{Chen2025} makes use of simultaneous start). Because of the strongly non Markovian nature of trail mediated interactions, it is expected that both protocols can yield different behaviours.  The second main result of this Letter is that, for $N$ SATW$_\phi$, the splitting probabilities are in fact identical in both protocols and can be written, for $0<k<N$,
\begin{align}
    &\p{+^k|x_0} = \binom{N+2(\phi-1)}{k+\phi-1} x_0^{k+\phi-1} (1-x_0)^{N-k+\phi-1}
    \label{eq:modified_binomial}\\
    &\p{+^N|x_0} = \p{+^0|1-x_0} = \frac{\int_0^{x_0} u^{N+\phi-2}(1-u)^{\phi-1} \mathrm{d}u}{\int_0^1 u^{N+\phi-2}(1-u)^{\phi-1}\mathrm{d}u},
    \label{eq:proba_all_same_wall}
\end{align}
where we made use of generalized binomial coefficients \footnote{Defined by $\binom{a}{b}=\frac{\Gamma(a+1)}{\Gamma(b+1)\Gamma(a-b+1)}$.} in \eqref{eq:modified_binomial} and we recognize the incomplete Beta function $I_{x_0}\left(N+\phi-1,\phi\right)$ in \eqref{eq:proba_all_same_wall}. Setting $\phi=1$ in Eqs.~\eqref{eq:modified_binomial}-\eqref{eq:proba_all_same_wall} recovers the expected binomial distribution, which is obtained straightforwardly for $N$ independent Brownian walkers. Although not obvious from Eqs.~\eqref{eq:modified_binomial}-\eqref{eq:proba_all_same_wall}, these probabilities are normalized i.e. $\sum_{k=0}^N \mathbb{P}(+^k|x_0) = 1$. These results are confirmed numerically Fig.~\ref{fig:panel_splittings}(a), and  illustrate further the nontrivial impact of trail mediated interactions on collective exploration. The main steps of the derivation are given below, and details can be found in SM. In contrast, our numerical analysis  for TSAWs unambiguously show that splitting probabilities can be protocol dependent (see Fig.~\ref{fig:panel_splittings}(b)).

\textit{Determination of the splitting probability $\p{+^0|x_0}$.} We first consider the event $(+^0 \, | \, x_0)$ where  $N$ SATW$_\phi$ random walkers, starting simultaneously from $x_0$, all eventually hit the target $0$ before $1$. The key step of our derivation relies on a generalization to $N$ SATW$_\phi$  random walkers of Ray-Knight theorems for SIRWs, originally developed for single random walkers by Toth et al. \cite{Toth_Generalized_RK_96}. We show in SM
that, in the scaling limit, the mutual local-time profile after absorption $L(x)$ admits the following exact representation
\begin{equation}
    L(x) \sim \begin{cases} \text{BESQ}^{2(N+\phi-1)}_{0}(x) \text{ if } x \in [0,x_0] \\
    \Tilde{\text{BESQ}}^{2(1-\phi)}_{L(x_0)}(x-x_0) \text{ if } x > x_0
    \end{cases}
    \label{eq:local_time_1_wall}
\end{equation}
where $\BESQ{\delta}_a$ (resp.\ $\BESQabs{\delta}_{a}$) denotes a squared Bessel process of dimension $\delta$ started from $a$ and reflected at 0 (resp.\ absorbed at 0) defined by the following stochastic differential equation~\cite{GoingJaeschkeYor2003}
\begin{equation}
    d\BESQ{\delta}(x) = \delta dx + 2\sqrt{\BESQ{\delta}(x)}dW_x
\end{equation}
where $dW_x$ is uncorrelated Gaussian white noise with variance 1. Crucially, this  representation allows for an exact integration of all realizations of trajectories contributing to the event  $(+^0\,|\,x_0)$. By noticing that  $L(x)$ defined in Eq.~\eqref{eq:local_time_1_wall} must hit $0$ for $x<1$ (otherwise at least one walker would have reached wall $1$), one obtains 
\begin{equation}
    \p{+^0|x_0} = \int_0^\infty \int_0^{1-x_0} P^{2(N + \phi-1)}_0(b,x_0) F^{2(1-\phi)}_0(b,x') \,db \,dx' 
    \label{eq:splitting_1_side}
\end{equation}
where $P^{\delta}_a(b,x)$ (resp.\ $F^{\delta}_0(a,x)$) is the transition density of BESQ$^\delta_a$ starting from $a$ to position $b$ (resp.\ the first-hitting density of 0) in time $x$~\cite{RevuzYor1991}. It is shown in SM that the integration of  Eq.~\eqref{eq:splitting_1_side} yields Eq.~\eqref{eq:proba_all_same_wall}.

\begin{figure}[ht]
    \centering
    \resizebox{\linewidth}{!}{\input{figures/local_time_N_2_SATW_one_wall.tikz}}
    \caption{One realization of the local time process for $2$ sequentially launched $\text{SATW}_\phi$ after both are absorbed at wall 0: in the space $[0,y]$ explored by the first particle (whose local time is in blue), the second particle is behaving as a simple symmetric and so is its local time (in red). The cumulative local time is in grey and is independent of the launching protocol, or of the speed of the walkers.~\label{fig:local_time_N_2_SATW_one_wall}}
\end{figure}
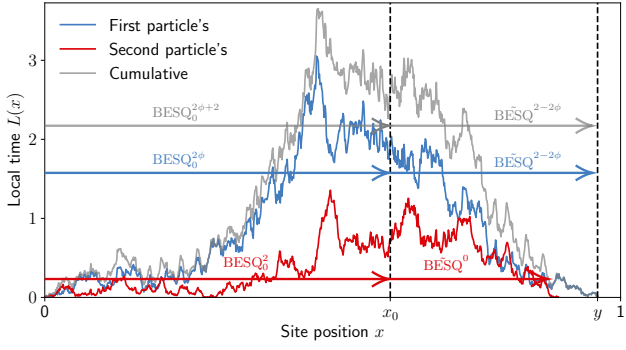

We now show that the exact representation of $L(x)$ for simultaneous starts given by Eq.~\eqref{eq:proba_all_same_wall} makes it possible to show that for SATW$_\phi$ random walkers, sequential and simultaneous starting conditions yield exactly the same splitting probabilities. For simplicity, we consider $N=2$, and the event $(+^0|x_0)$. For sequential starts, one can assume without loss of generality that the first walker $W_1$ has reached the right-most visited position $y<1$ upon hitting $0$ (see Fig.~\ref{fig:local_time_N_2_SATW_one_wall}): the corresponding local-time profile has been determined exactly (using similar Ray-Knight theorems for single SIRWs)  in~\cite{Bremont2025_exponents}, and showed to be represented as  $\BESQ{2\phi}_0$ from 0 to $x_0$ and a $\BESQabs{2-2\phi}$ from $x_0$ to $y$. After absorption of $W_1$, the second walker $W_2$ behaves as Brownian particle in $[0,y]$ (by definition of SATW$_\phi$),  and as a SATW$_\phi$ in $[y,1]$. As a consequence, its local time process in $[0,y]$ can be shown to be  $\BESQ{2}$ from $0$ to $x_0$,   $\BESQabs{0}$ between $x_0$ and $y$, and, if not yet absorbed, a $\BESQabs{2-2\phi}$ from $y$ to $1$ (see SM). This construction is shown Fig.~\ref{fig:local_time_N_2_SATW_one_wall}. The total local time is then the sum of the two profiles. Using the additivity of squared Bessel processes~\cite{GoingJaeschkeYor2003},
\begin{equation}
\BESQ{\delta_1}_{a_1} + \BESQ{\delta_2}_{a_2} \overset{\text{(law)}}{=} \BESQ{\delta_1 + \delta_2}_{a_1+a_2}
\end{equation}  
we recover Eq.~\eqref{eq:local_time_1_wall}, which shows that sequential and simultaneous starts are equivalent. This demonstration straightforwardly generalizes to  $N$ SATW$_\phi$ random walkers, and we show in SM that it also holds for any $(+^k \, | \, x_0)$ event. 
However, it holds only for the SATW$_\phi$ class, and in particular equivalence breaks down for TSAWs random walkers as illustrated Fig.~\ref{fig:panel_splittings}(b). We argue in SM that a simultaneous start yields an increased spreading of the distribution of the random walkers in space, and thus effectively enhances the effective repulsion between walkers.

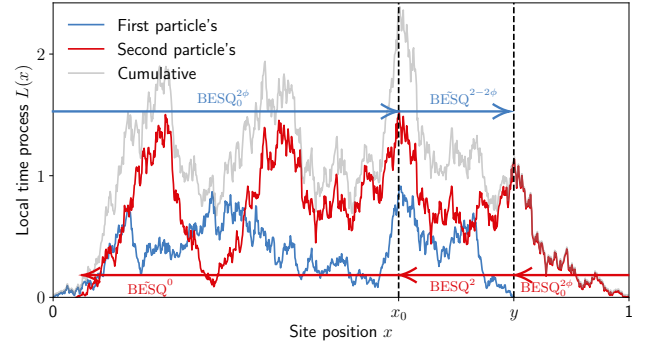
\begin{figure}[ht]
    \centering
    \resizebox{\linewidth}{!}{\input{figures/schema_2_SATW_2_walls.tikz}}
    \caption{One realization of the local time process for $2$ sequentially launched $\text{SATW}_\phi$ after the first (resp.\ second) one is absorbed at wall 0 (resp.\ one): in the space explored by the first particle, the second particle is behaving as a simple symmetric and so is its local time.\label{fig:schema_2_SATW_2_walls}}
\end{figure}

\textit{Determination of $\p{+^k \mid x_0}$ for $0<k<N$.} For generic $(+^k \, | \, x_0)$ events with $k>0$, it is more convenient to construct the local time profile for sequential starts, and make use of the equivalence given above for SATW$_\phi$ random walkers ; we consider  the case $k=1, N=2$ for simplicity.  First, we consider the event $(-+)$, where the first walker $W_1$ is absorbed at  0, and the second walker $W_2$ at  1. Suppose that $W_1$ has rightmost visited position $y<1$ upon  absorption: this occurs with probability $q_-^{\phi}(x_0,y)$, derived exactly e.g. in~\cite{Bremont2025_exponents} and the local time profile is the same as previous case (see Fig.~\ref{fig:schema_2_SATW_2_walls}). After absorption of $W_1$, walker $W_2$ starts its trajectory in the local time field $L^{(1)}(x)$ already deposited by $W_1$: the walker $W_2$ can thus be seen as a SATW$_\phi$ starting in an aged, already explored environment $[0,y]$. As was shown in \cite{chaumontUpperLower}, the local time of $W_2$ upon absorption at $1$ can be constructed as a  $\BESQ{2\phi}_0$ process running backward from $1$ to $y$, then a $\BESQ{2}$ from $y$ to $x_0$, and finally a $\BESQabs{0}$ from $x_0$ to absorption. This construction is shown in Fig.~\ref{fig:schema_2_SATW_2_walls}. Altogether, for a given a value of $y$, the probability $R(x_0,y)$ that $W_2$ is absorbed at  1 can be written
\begin{align}
    &R(x_0,y) = \label{eq:R(x_0,y)} \\
     &\int_0^{\infty}da \int_0^\infty db \int_0^{1-x_0}dz  P_0^{2\phi}(a,1-y) P_a^2(b,y-x_0) F_0^0(b,z). \nonumber
\end{align}
Averaging over $y$, we obtain the probability $\p{-+\mid x_0}$ that $W_1$ is absorbed at 0 and $W_2$ at 1:
\begin{align}
    \p{-+|x_0} &= \int_x^{1} q_{-}(x_0,y) R(x_0,y)\, dy \nonumber \\
    &= \frac{1}{2}\binom{2\phi}{\phi}\left(\left(1-x_0\right)x_0\right)^{\phi}.
    \label{eq:Pmp}
\end{align}
Summing the two equally likely scenarios $(-+)$ and $(+-)$ finally yields Eq.~\eqref{eq:modified_binomial} with $k=1, N=2$. This argument is generalized to   arbitrary $k,N$ in SM, and eventually shows Eq.~\eqref{eq:modified_binomial}.

\textit{N-walkers persistence exponents from splitting probabilities.}
We now generalize the approach of
\cite{Majumdar_hitting_persistence_2010,Levernier2018},
which relates the asymptotic behaviour of splitting probabilities
to persistence exponents, to the case of $N$ random walkers.
Consider $N$ walkers starting at $x_{i,0}>0$, for
$i=1,\dots,N$, on the semi-infinite line and absorbed at $0$.
Let $F_{k,N}(t\,|\,x_{i,0})$ be distribution of the $(N-k+1)^{\rm th}$ absorption time. We assume that
the trajectories are scale invariant, with walk dimension $d_w$
defined by $\langle x_i(t)^2\rangle\propto t^{2/d_w}$. Scale
invariance then yields
\begin{equation}
\label{scaling-fpt-kn}
    F_{k,N}(t\,|\,x_{i,0})
    =
    \frac{1}{t}
    f_{k,N}\left(
        \frac{x_{1,0}^{d_w}}{t},
        \dots,
        \frac{x_{N,0}^{d_w}}{t}
    \right).
\end{equation}

We now scale the starting positions as
$x_{i,0}=z\widetilde{x}_{i,0}$, where $z\to0$ and the
$\widetilde{x}_{i,0}$ are kept fixed. By definition of the
generalized persistence exponent, we have
$F_{k,N}(t\,|\,x_{i,0})\propto t^{-1-\theta_{k,N}}$ at long
times. Equation~\eqref{scaling-fpt-kn} therefore implies
\begin{equation}
    f_{k,N}\left(
        \frac{z^{d_w}\widetilde{x}_{1,0}^{d_w}}{t},
        \dots,
        \frac{z^{d_w}\widetilde{x}_{N,0}^{d_w}}{t}
    \right)
    \underset{z\to0}{\propto}
    z^{d_w\theta_{k,N}}t^{-\theta_{k,N}}.
\end{equation}

We now consider the same walkers in the interval $[0,1]$, and denote by $Q_{k,N}(x_{i,0})$ the probability that at least $k$ of them are absorbed at $1$. The walkers reach $0$ within typical timescales $x_{i,0}^{d_w}\propto z^{d_w}$, which, as $z\to0$, are much smaller than the timescale $(1-z)^{\dw}$ of order unity needed to reach the opposite boundary. Thus, the probability that at least $k$ walkers reach $1$ is determined mostly by trajectories for which the $(N-k+1)^{\rm th}$ absorption at $0$ occurs only after an atypically large time $t$ of order unity. Consequently,
\begin{align}
    Q_{k,N}(z\widetilde{x}_{i,0})
    \underset{z\to0}{\propto}
    \int_1^\infty
    F_{k,N}(t\,|\,z\widetilde{x}_{i,0})\,dt \underset{z\to0}{\propto}
    z^{d_w\theta_{k,N}}.
    \label{eq:decay_Q}
\end{align}

For the SATW$_\phi$, when all walkers start from the same position
$x_0$, the splitting probabilities $Q_{k,N}$ are given by
Eqs.~\eqref{eq:modified_binomial}
and~\eqref{eq:proba_all_same_wall}, which yields the asymptotics
\begin{equation}
    Q_{k,N}(x_0)
    =
    \sum_{l=k}^{N}\p{+^l|x_0}
    \underset{x_0\to0}{\propto}
    x_0^{k+\phi-1}.
\end{equation}
Making use of the walk dimension $d_w=2$ for
SATW$_\phi$~\cite{Toth_Generalized_RK_96}, we deduce from
Eq.~\eqref{eq:decay_Q} the exact values of the
persistence exponents $\theta_{k,N}$ for SATW$_\phi$, given in
Eq.~\eqref{eq:theta_kN}.

\textit{Determination of $\theta_{1,N}^{(\text{SIRWs})}$.} Importantly, the exact results presented so far have been obtained for the SATW$_\phi$ class, by making use of explicit constructions of the local time profile. As of now, such constructions are not available for other classes of SIRWs. However, the persistence exponent $\theta_{1,N}^{(SIRW)}$ can be obtained by direct arguments for all SIRWs as we show now.
Consider  $N$  SIRWs on the semi-infinite space with absorbing boundary. We look for the persistence exponent $\theta_{1,N}^{(SIRW)}$ of the last walker alive. We argue that, in the limit $t \to\infty$, after all the other $N-1$ walkers have been absorbed, the time spent by the last  walker $W$ in the  interval $\mathcal{R}$ visited by at least one other walker becomes negligible compared to the time spent in unexplored areas. Indeed, the probability for $W$ to stay for a time $t$ in the finite interval $\mathcal{R}$ decays exponentially with $t$. In turn, the survival probability outside $\mathcal{R}$  decays as $t^{-\theta_{1,1}^{\rm (SIRW)}}$ because $W$ is free from the influence of the other walkers  outside of $\mathcal{R}$. Finally, this yields $\theta_{1,N}^{\rm (SIRW)} = \theta_{1,1}^{\rm (SIRW)}$, as is confirmed numerically in Fig.~\ref{fig:panel_fits_FPT}(b).

\textit{Discussion and conclusion.} We have established an exact analytical framework for collective first-passage observables of trail-interacting random walkers. For the broad class of saturating self-interacting walks, we obtained the complete spectrum of generalized persistence exponents and exact splitting probabilities for arbitrary numbers of walkers. Our analysis  is based on an exact representation of the collective trail field in terms of squared Bessel processes.
A particularly striking result is the equivalence between sequential and simultaneous exploration protocols -- both of potential experimental relevance -- for saturating trail interactions. Although walkers interact through a common memory field and therefore exhibit strong history dependence, the statistics of competing targets remain exactly unchanged. This property contrasts sharply with the behavior of non-saturating interactions, for which our numerical results demonstrate that the two protocols become genuinely different. 
More generally, our work provides quantitative predictions for first-passage observables that are in principle accessible experimentally in biological or synthetic active systems interacting through persistent environmental memory, such as active droplets: trail-mediated interactions have recently been observed to enhance the spatial spreading of active droplets depositing persistent trails compared with isolated particles  with a simultaneous start protocol \cite{Chen2025}, an effect qualitatively consistent with our results.

\textit{Author contributions.}
P.P. and J.B. developed the theory and performed the analytical calculations. P.P. conducted the numerical simulations and analysis. P.P., J.B., O.B. and  R.V. jointly wrote the manuscript. P.P. wrote the Supplementary Material.


\bibliography{biblio}

\newpage

\insertpdfpage{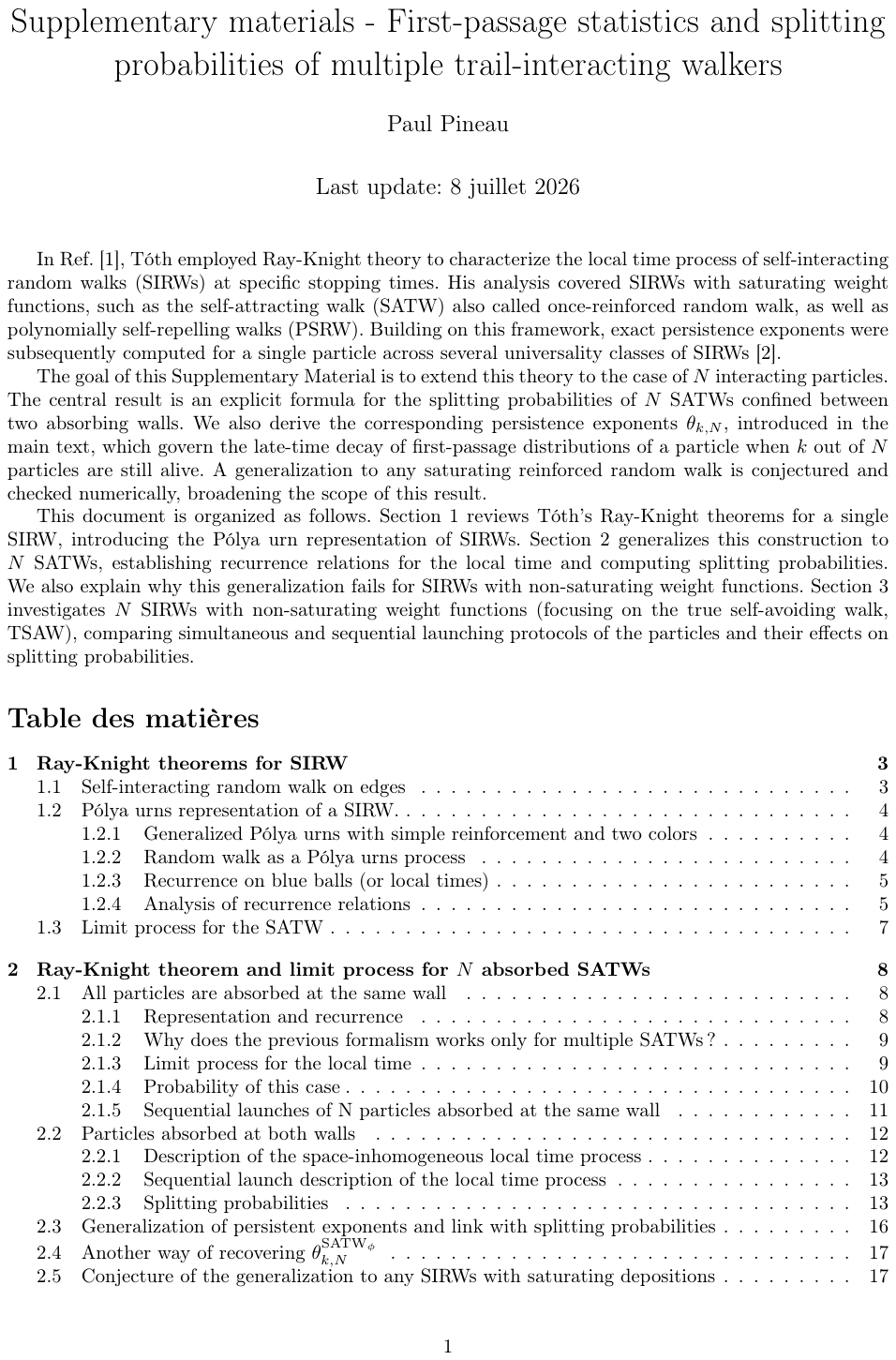}{1}
\insertpdfpage{Supplementary.pdf}{2}
\insertpdfpage{Supplementary.pdf}{3}
\insertpdfpage{Supplementary.pdf}{4}
\insertpdfpage{Supplementary.pdf}{5}
\insertpdfpage{Supplementary.pdf}{6}
\insertpdfpage{Supplementary.pdf}{7}
\insertpdfpage{Supplementary.pdf}{8}
\insertpdfpage{Supplementary.pdf}{9}
\insertpdfpage{Supplementary.pdf}{10}
\insertpdfpage{Supplementary.pdf}{11}
\insertpdfpage{Supplementary.pdf}{12}
\insertpdfpage{Supplementary.pdf}{13}
\insertpdfpage{Supplementary.pdf}{14}
\insertpdfpage{Supplementary.pdf}{15}
\insertpdfpage{Supplementary.pdf}{16}
\insertpdfpage{Supplementary.pdf}{17}
\insertpdfpage{Supplementary.pdf}{18}
\insertpdfpage{Supplementary.pdf}{19}
\insertpdfpage{Supplementary.pdf}{20}
\insertpdfpage{Supplementary.pdf}{21}
\insertpdfpage{Supplementary.pdf}{22}
\insertpdfpage{Supplementary.pdf}{23}
\insertpdfpage{Supplementary.pdf}{24}
\insertpdfpage{Supplementary.pdf}{25}
\insertpdfpage{Supplementary.pdf}{26}
\insertpdfpage{Supplementary.pdf}{27}
\insertpdfpage{Supplementary.pdf}{28}

\end{document}

%% file: figures/schema.tikz
\tikzset{every picture/.style={line width=0.75pt}} 
\pgfdeclarelayer{backmost}
\pgfsetlayers{backmost,background,main}

\begin{tikzpicture}[x=0.75pt,y=0.75pt,yscale=-1,xscale=1]

\begin{pgfonlayer}{backmost}
\draw  [color= blueblind ,draw opacity=1 ][fill=blueblind  ,fill opacity=0.2 ] (29,122.4) .. controls (29,121.07) and (30.07,120) .. (31.4,120) -- (208.6,120) .. controls (209.93,120) and (211,121.07) .. (211,122.4) -- (211,129.6) .. controls (211,130.93) and (209.93,132) .. (208.6,132) -- (31.4,132) .. controls (30.07,132) and (29,130.93) .. (29,129.6) -- cycle ;

\draw  [color=blueblind ,draw opacity=1 ][fill=blueblind  ,fill opacity=0.2 ] (212,86) .. controls (213.55,86) and (214.8,87.25) .. (214.8,88.8) -- (214.8,97.2) .. controls (214.8,98.75) and (213.55,100) .. (212,100) -- (159,100) .. controls (159,100) and (159,100) .. (159,100) -- (159,86) .. controls (159,86) and (159,86) .. (159,86) -- cycle ;

\draw   [color=blueblind ,draw opacity=1 ][fill=blueblind  ,fill opacity=0.2 ]  (19,86) -- (122,86) -- (122,100) -- (19,100) -- cycle ;
\end{pgfonlayer}

\draw    (280,126) -- (233,126) -- (0,126) (270,128) -- (270,124)(260,128) -- (260,124)(250,128) -- (250,124)(240,128) -- (240,124)(230,128) -- (230,124)(220,128) -- (220,124)(210,128) -- (210,124)(200,128) -- (200,124)(190,128) -- (190,124)(180,128) -- (180,124)(170,128) -- (170,124)(160,128) -- (160,124)(150,128) -- (150,124)(140,128) -- (140,124)(130,128) -- (130,124)(120,128) -- (120,124)(110,128) -- (110,124)(100,128) -- (100,124)(90,128) -- (90,124)(80,128) -- (80,124)(70,128) -- (70,124)(60,128) -- (60,124)(50,128) -- (50,124)(40,128) -- (40,124)(30,128) -- (30,124)(20,128) -- (20,124)(10,128) -- (10,124) ;

\draw   [very thick] (0,121) -- (0,131)   node [below]   {$0$}; 
\draw   [very thick] (280,121) -- (280,131)   node [below]   {$1$};


\draw   (14,61.65) .. controls (14,55.78) and (18.76,51) .. (24.63,51) -- (115.37,51) .. controls (121.24,51) and (126,55.78) .. (126,61.65) -- (126,93.55) .. controls (126,99.42) and (121.24,104.18) .. (115.37,104.18) -- (24.63,104.18) .. controls (18.76,104.18) and (14,99.42) .. (14,93.55) -- cycle ;
\draw    (27.63,104.18) -- (56,119.42) ;
\draw   (56,119.42) .. controls (56,116.99) and (57.97,115) .. (60.4,115) -- (79.6,115) .. controls (82.03,115) and (84,116.99) .. (84,119.42) -- (84,132.62) .. controls (84,135.05) and (82.03,137) .. (79.6,137) -- (60.4,137) .. controls (57.97,137) and (56,135.05) .. (56,132.62) -- cycle ;
\draw    (115.37,104.18) -- (84,119.42) ;
\draw    (19,93) -- (122,93) (33,89) -- (33,97)(73,89) -- (73,97)(113,89) -- (113,97) ;
\draw  [color=redblind][line width=0.75]   (76,84) .. controls (89.37,70.81) and (100,74.77) .. (109.64,82.75) ;
\draw [shift={(111,84)}, rotate = 220] [color=redblind][line width=0.75]   (10.93,-3.29) .. controls (6.95,-1.4) and (3.31,-0.3) .. (0,0) .. controls (3.31,0.3) and (6.95,1.4) .. (10.93,3.29)   ;
\draw   [color=redblind][line width=0.75]  (70,84) .. controls (58.48,70.74) and (48.8,72.97) .. (36.55,82.75) ;
\draw   [shift={(35,84)}, rotate = 320] [color=redblind][line width=0.75]   (10.93,-3.29) .. controls (6.95,-1.4) and (3.31,-0.3) .. (0,0) .. controls (3.31,0.3) and (6.95,1.4) .. (10.93,3.29)   ;
\draw    (150,121) -- (150,131) ;
\draw (144,135) node [anchor=north west][inner sep=0.75pt]    {\small{$x_{0}$}};
\draw   (154,61.65) .. controls (154,55.78) and (158.76,51) .. (164.63,51) -- (255.37,51) .. controls (261.24,51) and (266,55.78) .. (266,61.65) -- (266,93.55) .. controls (266,99.42) and (261.24,104.18) .. (255.37,104.18) -- (164.63,104.18) .. controls (158.76,104.18) and (154,99.42) .. (154,93.55) -- cycle ;
\draw    (167.63,104.18) -- (196,119.42) ;
\draw   (196,119.42) .. controls (196,116.99) and (197.97,115) .. (200.4,115) -- (219.6,115) .. controls (222.03,115) and (224,116.99) .. (224,119.42) -- (224,132.62) .. controls (224,135.05) and (222.03,137) .. (219.6,137) -- (200.4,137) .. controls (197.97,137) and (196,135.05) .. (196,132.62) -- cycle ;
\draw    (255.37,104.18) -- (224,119.42) ;
\draw    (185,93) -- (159,93) -- (263,93) (173,89) -- (173,97)(213,89) -- (213,97)(253,89) -- (253,97) ;
\draw  [color=redblind][line width=0.75]  (216,84) .. controls (229.37,70.81) and (240,74.77) .. (249.64,82.75) ;
\draw [shift={(251,84)}, rotate = 220] [color=redblind][line width=0.75]    (10.93,-3.29) .. controls (6.95,-1.4) and (3.31,-0.3) .. (0,0) .. controls (3.31,0.3) and (6.95,1.4) .. (10.93,3.29)   ;
\draw [color=redblind][line width=0.75]   (210,84) .. controls (198.48,70.74) and (188.8,72.97) .. (176.55,82.75) ;
\draw [shift={(175,84)}, rotate = 320.19] [color=redblind  ][line width=0.75]    (10.93,-3.29) .. controls (6.95,-1.4) and (3.31,-0.3) .. (0,0) .. controls (3.31,0.3) and (6.95,1.4) .. (10.93,3.29)   ;

\begin{scope}[shift={(0,-18)}]  
\draw [color = greenblind]   (160,165) .. controls (198.55,181.29) and (245.05,184.25) .. (272.56,165.86)  node[midway, above] {$k$};
\draw [shift={(273.8,165)}, rotate = 144.73] [color=greenblind ][line width=0.75]    (10.93,-3.29) .. controls (6.95,-1.4) and (3.31,-0.3) .. (0,0) .. controls (3.31,0.3) and (6.95,1.4) .. (10.93,3.29)   ;
\draw [color = greenblind]   (140,165) .. controls (104.72,181.99) and (29.1,181.69) .. (6.62,165.98) node[midway, above] {$N-k$} ;
\draw [shift={(5,165)}, rotate = 32.66] [color=greenblind][line width=0.75]    (10.93,-3.29) .. controls (6.95,-1.4) and (3.31,-0.3) .. (0,0) .. controls (3.31,0.3) and (6.95,1.4) .. (10.93,3.29)   ;
\draw [color = greenblind] (130,175) node [anchor=north west][inner sep=0.75pt]    {$\boldsymbol{(+^k|x_0)}$};
\end{scope}

\draw (45,52) node [anchor=north west][inner sep=0.75pt]    {$\frac{1}{2}$};
\draw (87,52) node [anchor=north west][inner sep=0.75pt]    {$\frac{1}{2}$};
\draw (179,52) node [anchor=north west][inner sep=0.75pt]    {$\frac{\phi}{1+\phi}$};
\draw (221,52) node [anchor=north west][inner sep=0.75pt]    {$\frac{1}{1+\phi}$};
\begin{scope}[on background layer]
\draw[fill=redblind, draw=none] (70,126) circle (4);
\draw[fill=redblind, draw=none] (90,126) circle (4);
\draw[fill=redblind, draw=none] (210,126) circle (4);
\draw[fill=redblind, draw=none] (73,93) circle (6);
\draw[fill=redblind, draw=none] (213,93) circle (6);
\end{scope}

\end{tikzpicture}

%% file: figures/local_time_N_2_SATW_one_wall.tikz
\tikzset{every picture/.style={line width=0.75pt}} 
\begin{tikzpicture}[x=0.75pt,y=0.75pt,yscale=-1,xscale=1]

\draw (330.16,148.99) node  {\includegraphics[width=403.74pt,height=224.98pt]{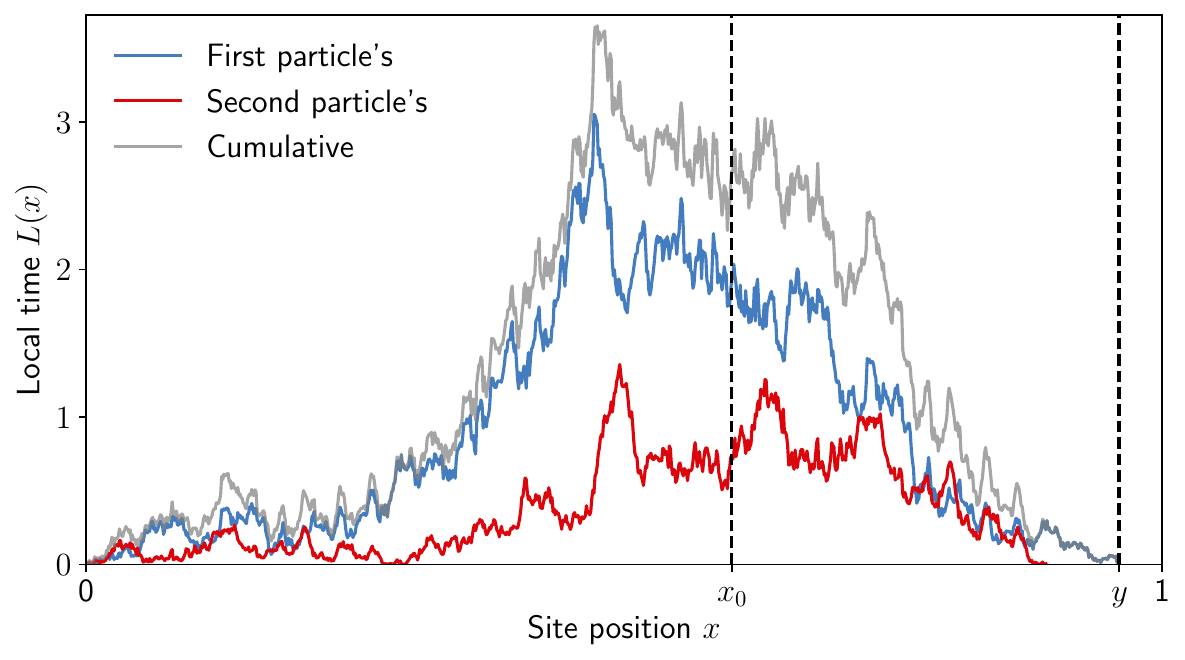}};
\draw [color=blueblind  ,draw opacity=1 ][line width=1.5]    (100,150) -- (392,150) ;
\draw [shift={(392,150)}, rotate = 179.6] [color=blueblind  ,draw opacity=1 ][line width=1.5]    (14.21,-6.37) .. controls (9.04,-2.99) and (4.3,-0.87) .. (0,0) .. controls (4.3,0.87) and (9.04,2.99) .. (14.21,6.37)   ;
\draw [color=blueblind  ,draw opacity=1 ][line width=1.5]    (392,150) -- (565, 150) ;
\draw [shift={(565,150)}, rotate = 180] [color=blueblind  ,draw opacity=1 ][line width=1.5]    (14.21,-6.37) .. controls (9.04,-2.99) and (4.3,-0.87) .. (0,0) .. controls (4.3,0.87) and (9.04,2.99) .. (14.21,6.37)   ;
\draw [color=redblind  ,draw opacity=1 ][line width=1.5]    (100,240) -- (392,240) ;
\draw [shift={(392,240)}, rotate = 179.6] [color=redblind ,draw opacity=1 ][line width=1.5]    (14.21,-6.37) .. controls (9.04,-2.99) and (4.3,-0.87) .. (0,0) .. controls (4.3,0.87) and (9.04,2.99) .. (14.21,6.37)   ;
\draw [color=redblind  ,draw opacity=1 ][line width=1.5]    (392,240) -- (526,240) ;
\draw [shift={(527,240)}, rotate = 180] [color={rgb, 255:redblind, 255; green, 0; blueblind, 0 }  ,draw opacity=1 ][line width=1.5]    (14.21,-6.37) .. controls (9.04,-2.99) and (4.3,-0.87) .. (0,0) .. controls (4.3,0.87) and (9.04,2.99) .. (14.21,6.37)   ;
\draw (190,130) node [anchor=north west][inner sep=0.75pt]  [color=blueblind  ,opacity=1 ]  [align=left] {$\BESQ{2\phi}_0$};
\draw (480,130) node [anchor=north west][inner sep=0.75pt]  [color=blueblind  ,opacity=1 ]  [align=left] {$\BESQabs{2-2\phi}$};
\draw (420,218) node [anchor=north west][inner sep=0.75pt]  [color=redblind  ,opacity=1 ] [align=left] {$\BESQabs{0}$};
\draw (250,218) node [anchor=north west][inner sep=0.75pt]  [color=redblind ,opacity=1 ] [align=left] {$\BESQ{2}_0$};

\draw [color={rgb, 1: red,0.5019607843137255; green, 0.5019607843137255; blue, 0.5019607843137255}  ,draw opacity=0.7 ][line width=1.5]    (100,110) -- (392,110) ;
\draw [shift={(392,110)}, rotate = 179.6] [color={rgb, 1: red,0.5019607843137255; green, 0.5019607843137255; blue, 0.5019607843137255}  ,draw opacity=1 ][line width=1.5]    (14.21,-6.37) .. controls (9.04,-2.99) and (4.3,-0.87) .. (0,0) .. controls (4.3,0.87) and (9.04,2.99) .. (14.21,6.37)   ;
\draw [color={rgb, 1: red,0.5019607843137255; green, 0.5019607843137255; blue, 0.5019607843137255}  ,draw opacity=0.7 ][line width=1.5]    (392,110) -- (565,110) ;
\draw [shift={(565,110)}, rotate = 180] [color={rgb, 1: red,0.5019607843137255; green, 0.5019607843137255; blue, 0.5019607843137255}  ,draw opacity=0.7 ][line width=1.5]    (14.21,-6.37) .. controls (9.04,-2.99) and (4.3,-0.87) .. (0,0) .. controls (4.3,0.87) and (9.04,2.99) .. (14.21,6.37)   ;
\draw (190,90) node [anchor=north west][inner sep=0.75pt]  [color={rgb, 1: red,0.5019607843137255; green, 0.5019607843137255; blue, 0.5019607843137255}  ,opacity=1 ]  [align=left] {$\BESQ{2\phi + 2}_0$};
\draw (480,90) node [anchor=north west][inner sep=0.75pt]  [color={rgb, 1: red,0.5019607843137255; green, 0.5019607843137255; blue, 0.5019607843137255}  ,opacity=1 ]  [align=left] {$\BESQabs{2-2\phi}$};

\end{tikzpicture}

%% file: figures/schema_2_SATW_2_walls.tikz
\tikzset{every picture/.style={line width=0.75pt}} 
\begin{tikzpicture}[x=0.75pt,y=0.75pt,yscale=-1,xscale=1]

\draw (329.55,150.17) node  {\includegraphics[width=404.32pt,height=225.25pt]{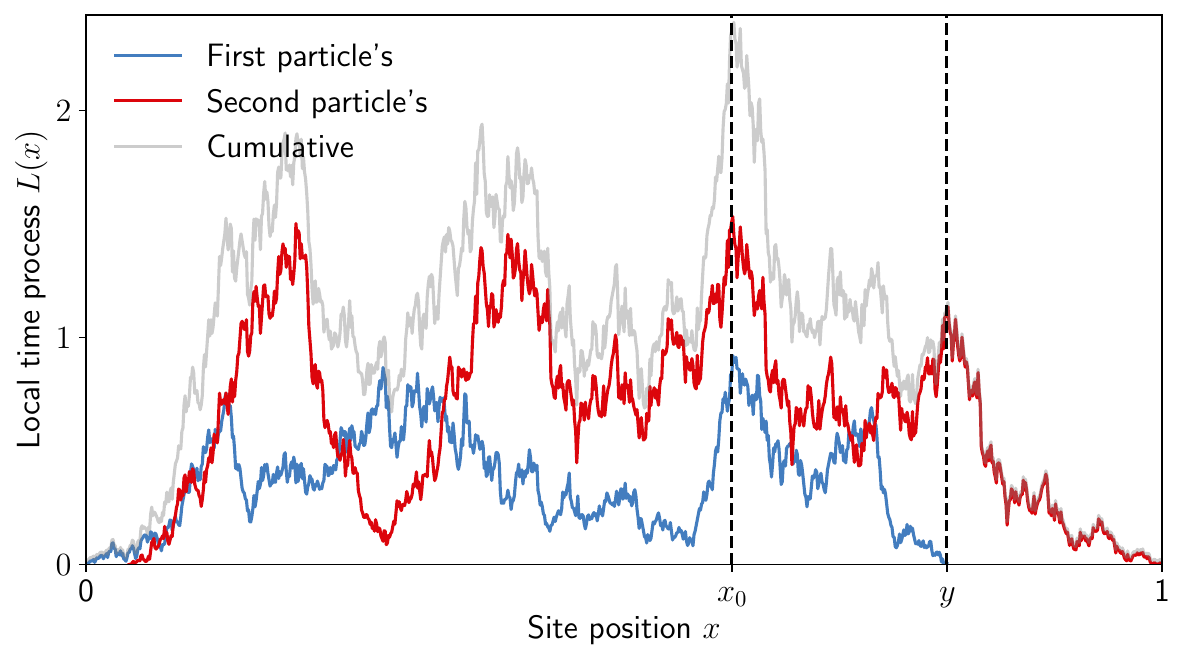}};

\begin{scope}[shift={(0,99)}]
    \draw [color=blueblind  ,draw opacity=1 ][line width=1.5]    (98,0) -- (389,0) ;
    \draw [shift={(392,0)}, rotate = 180] [color=blueblind  ,draw opacity=1 ][line width=1.5]    (14.21,-6.37) .. controls (9.04,-2.99) and (4.3,-0.87) .. (0,0) .. controls (4.3,0.87) and (9.04,2.99) .. (14.21,6.37)   ;
    \draw [color={rgb, 255:redblind, 0; green, 0; blueblind, 255 }  ,draw opacity=1 ][line width=1.5]    (392,0) -- (485,0) ;
    \draw [shift={(487,0)}, rotate = 180] [color=blueblind  ,draw opacity=1 ][line width=1.5]    (14.21,-6.37) .. controls (9.04,-2.99) and (4.3,-0.87) .. (0,0) .. controls (4.3,0.87) and (9.04,2.99) .. (14.21,6.37)   ;

    \draw (220,-20) node [anchor=north west][inner sep=0.75pt]    [color=blueblind  ,opacity=1 ] {$\BESQ{2\phi}_0$};
    \draw (418,-21) node [anchor=north west][inner sep=0.75pt]     [color=blueblind  ,opacity=1 ]{$\BESQabs{2-2\phi}$};

\end{scope}

\begin{scope}[shift={(0,240)}]
    \draw [color=redblind  ,draw opacity=1 ][line width=1.5]    (589,-2) -- (495,-2) ;
    \draw [shift={(492,-2)}, rotate = 0] [color=redblind  ,draw opacity=1 ][line width=1.5]    (14.21,-6.37) .. controls (9.04,-2.99) and (4.3,-0.87) .. (0,0) .. controls (4.3,0.87) and (9.04,2.99) .. (14.21,6.37)   ;
    \draw [color=redblind  ,draw opacity=1 ][line width=1.5]    (492,-2) -- (397,-2) ;
    \draw [shift={(394,-2)}, rotate = 0] [color=redblind  ,draw opacity=1 ][line width=1.5]    (14.21,-6.37) .. controls (9.04,-2.99) and (4.3,-0.87) .. (0,0) .. controls (4.3,0.87) and (9.04,2.99) .. (14.21,6.37)   ;
    \draw [color=redblind  ,draw opacity=1 ][line width=1.5]    (394,-2) -- (123,-2) ;
    \draw [shift={(125,-2)}, rotate = 0] [color=redblind  ,draw opacity=1 ][line width=1.5]    (14.21,-6.37) .. controls (9.04,-2.99) and (4.3,-0.87) .. (0,0) .. controls (4.3,0.87) and (9.04,2.99) .. (14.21,6.37)   ;

    \draw (161,0) node [anchor=north west][inner sep=0.75pt]  [color=redblind  ,opacity=1 ]  {$\BESQabs{0}$};
    \draw (418,0) node [anchor=north west][inner sep=0.75pt]  [color=redblind  ,opacity=1 ]  {$\BESQ{2}$};
    \draw (495,0) node [anchor=north west][inner sep=0.75pt]  [color=redblind  ,opacity=1 ]  {$\BESQ{2\phi}_0$};
\end{scope}

\end{tikzpicture}